\documentclass[journal]{IEEEtran}
\usepackage{cite}
\usepackage{graphicx,amssymb,lineno}
\usepackage{amsmath,amsfonts,amssymb}
\newtheorem{theorem}{Theorem}

\usepackage{algorithm}
\usepackage{algorithmic}
\usepackage[usenames]{color}
\usepackage{float}
\usepackage{mathtools}
\usepackage{cite}
\usepackage{bm}

\usepackage{graphicx,graphics,color,epsfig,subfigure,graphpap,rotate}
\usepackage{times, verbatim, subfigure, epsfig, graphicx, latexsym, amsmath}
\usepackage{url}
\usepackage{subfigure}
\usepackage{CJK}

\begin{document}

\title{Reconfigurable Intelligent Surface Assisted Device-to-Device Communications}

\author{Yali~Chen,
        Bo~Ai,~\IEEEmembership{Senior Member,~IEEE},
        Hongliang~Zhang,~\IEEEmembership{Member,~IEEE},
        Yong~Niu,~\IEEEmembership{Member,~IEEE},
        Lingyang~Song,~\IEEEmembership{Fellow,~IEEE},
        Zhu~Han,~\IEEEmembership{Fellow,~IEEE},
        and H. Vincent Poor,~\IEEEmembership{Fellow,~IEEE}

\thanks{Y. Chen, B. Ai, and Y. Niu are with the State Key Laboratory of Rail Traffic Control and Safety, Beijing Engineering Research Center of High-speed Railway Broadband Mobile Communications, and the School of Electronic and Information Engineering, Beijing Jiaotong University, Beijing 100044, China (e-mails: chenyali@bjtu.edu.cn; boai@bjtu.edu.cn; niuy11@163.com).}

\thanks{H. Zhang is with the Department of Electrical and Computer Engineering, University of Houston, Houston, TX 77004 USA (e-mail: hongliang.zhang92@gmail.com).}

\thanks{L. Song is with the Department of Electronics Engineering, Peking University, Beijing, China (e-mail: lingyang.song@pku.edu.cn).}

\thanks{Z. Han is with the Department of Electrical and Computer Engineering,
University of Houston, Houston, TX 77004 USA (e-mail: zhan2@uh.edu).}

\thanks{H. Vincent Poor is with School of Engineering and Applied Science, Princeton University, NY, USA (e-mail: poor@princeton.edu).}

}

\maketitle

\begin{abstract}
With the evolution of the 5G, 6G and beyond, device-to-device (D2D) communication has been developed as an energy-, and spectrum-efficient solution. In cellular network, D2D links need to share the same spectrum resources with the cellular link. A reconfigurable intelligent surface (RIS) can reconfigure the phase shifts of elements and create favorable beam steering, which can mitigate aggravated interference caused
by D2D links. In this paper, we study a RIS-assisted single cell uplink communication network scenario, where the cellular link and multiple D2D links utilize direct propagation and reflecting one-hop propagation. The problem of maximizing the total system rate is formulated by jointly optimizing transmission powers of all links and discrete phase shifts of all elements. The formulated problem is an NP-hard mixed integer non-convex non-linear problem. To obtain practical solutions, we capitalize on alternating maximization and the problem is decomposed into
two sub-problems. For the power allocation, the problem is a difference of concave functions (DC) problem, which is solved with the gradient descent method. For the phase shift, a local search algorithm with lower complexity is utilized. Simulation results show that deploying RIS and optimizing the phase shifts have a significant effect on mitigating D2D network interference.
\end{abstract}

\section{Introduction}\label{S1}

In recent years, with the popularity of mobile devices and smart terminals, the demand for data traffic in wireless networks has increased dramatically. In order to meet this rapid growth of data traffic and achieve seamless communication, device-to-device (D2D) technology is considered to be a promising technique \cite{zhao}. In D2D communications, users that are physically close to each other can communicate directly without forwarding through a base station (BS). Due to the short distance transmission, D2D communications can reduce energy consumption, provide early warning in emergencies, and upgrade users' quality of service (QoS) demands \cite{kim,chenyali,song,choi}. Typically, D2D is allowed to share the uplink spectrum with cellular users, which overcomes the shortage of spectrum. However, the inevitable interference requires to protect cellular networks from harmful interferences under the strict requirement of communication quality \cite{gu}. Recently, an innovative and revolutionary technology is developed, namely reconfigurable intelligent surface (RIS), and dedicated to its role in beyond 5G networks \cite{shen}. RIS can be configured to control the reflection, refraction and scattering of electromagnetic waves to shape the channel, which can be utilized to effectively eliminate D2D interference and fulfill demanding data rates.

RIS is a two-dimensional ultra-thin reflecting surface with integrated electronic circuits \cite{wu1,RIS1,RIS3,hongliang}. It consists of a number of elements that are controlled by programmable varactor diodes. For each element, electromagnetic response, i.e., phase shift towards incident signals, can be tuned in a software-defined way. With the reconfigurable property, the incident signals are superposed and reflected by adjusting phase shifts to create a favorable beam steering towards user, in order to effectively control the multi-path effects \cite{xu}. Different from traditional amplify-and-forward (AF), backscatter communication and massive multiple input multiple output (MIMO), RIS works by forwarding a more efficient phase-shifted version of the signals and shaping channel propagation to adapt against channel variations due to harsh and unpredictable wireless environments. The array architecture is passive and its operating mechanism is reflecting rather than regenerating. Thus, it is battery-less with no radio frequency encoding, decoding and power consumption \cite{basar1}. To sum up, the function of RIS is equivalent to fine-grained 3D passive beamforming. By adjusting the amplitude and/or phase shift of RIS element, the purpose of boosting the received intended power gain and destructively mitigating interference can be achieved, and consequently the improvements in spectrum and energy efficiency are obtained \cite{basar2}.

In general, extended coverage, improved transmission rate and reduced energy cost are achieved with RIS. As an artificial structure, RIS has broad application prospects \cite{RIS4}, such as cognitive radio network, wireless power transfer, and so on. For D2D communications, collaborating with RIS is very challenging and research on this topic has not yet been fully carried out. In addition to direct path, the RIS-assisted one-hop reflective path carries useful signals. The signals of both paths are superimposed at the receivers to achieve amplification gain and eliminate interference as much as possible. Besides, extremely small hardware footprint of RIS allows it to be easily attached to walls, building facades or interior ceiling. Flexible replacement, deployment and scalable cost make it convenient to be integrated into existing networks \cite{RIS2}.

In this paper, we study an uplink RIS-assisted heterogeneous network where a cellular user shares the same frequency band with multiple D2D links. The goal is to maximize the system transmission rate subjected to the QoS constraints and transmission power budget. Optimization variables include limited discrete phase shifts and continuous transmission power, both of which are coupled and not independent in the numerator and denominator of the signal to interference plus noise ratio (SINR) term in the objective function and constraints. The problem is a mixed integer non-convex non-linear problem, and really challenging to solve. In particular, due to co-channel interference and the fact that D2D transmitters are not capable of precoding, the optimal scheme is very hard to obtain \cite{lingyang}. Thus, we decompose the optimization problem into two sub-problems and solve iteratively.

The contributions of this paper are summarized as follows.

\begin{itemize}
\item We establish an uplink RIS-aided heterogeneous network, in which one cellular link and multiple D2D links are transmitted through a reflective path supported by RIS and a standard direct path to enhance the receiving gain. Based on the system model, we study transmission power design and discrete phase shift reconfiguration for all RIS elements.

\item We formulate the optimization problem of maximizing the system sum rate into a mixed integer non-convex non-linear problem. The minimum SINR and frequency response phase shift constraints should be satisfied. Since the optimization parameters are not independent, the closed-form solution is extremely difficult to get. Then, we use alternating iteration to update two separated sub-problems, power allocation and discrete phase shift optimizations.

\item The power allocation sub-problem is solved by difference of concave/convex functions (DC) programming theory \cite{DC}, multivariate Taylor expansion linearization, and classic gradient descent method. Another computationally affordable approach, namely local search, is adopted to optimize the phase shift.

\item Under the RIS characteristic parameters and other network parameters, we compare the impact of the proposed scheme, the scheme without the assistance of RIS and other typical schemes on the system performance. Finally, we prove that the interference of D2D networks with RIS deployed and phase shifts optimization is decreased significantly and the communication quality is greatly improved.

\end{itemize}

The rest of the paper is organized as follows. In Section~\ref{S2}, we summarize some related works. In Section~\ref{S3}, we establish the system model, including the description of the investigated system and the adopted channel models. Then, we formulate the sum rate maximization problem and decompose it into a power allocation sub-problem and a discrete phase shift optimization sub-problem in Section~\ref{S4}. Next, two sub-problem algorithms are designed, and the sum rate maximization algorithm is obtained by alternately iterating the two sub-problems in Section~\ref{S5}. Meanwhile, the theoretical analysis of the proposed algorithm is also provided. In Section~\ref{S6}, we conduct a performance evaluation and compare the proposed algorithm with other representative algorithms. Finally, we conclude this paper in Section~\ref{S7}.

\section{Related Work}\label{S2}
There have been several related works studying D2D network interference. Yu \emph{et al.} \cite{yuguanding} proposed a scheme that combined power control, channel assignment and mode selection to solve the interference problem in D2D communications, and further maximize the system throughput with the SINR constraints of all links satisfied. Available modes included reuse mode, dedicated mode and cellular mode. Wang \emph{et al.} \cite{wang} proposed a non-cooperative game-theoretic approach to allocate radio resources for D2D users with the optimization goal of maximizing the total battery lifetime. Besides, pricing was added as penalty for game players who selfishly occupied resources. These works do not change the channel environment, and then alleviate the D2D interference just by proper allocation.

There are many related works about RIS assisted wireless communications. In an uplink RIS-assisted communication scenario, Zhang \emph{et al.} \cite{zhang} analyzed the transmission rate in the case of continuous phase shifts for RIS configuration and focused on the effect of the number of discrete phase shifts on obtaining equal transmission rate in order to improve the QoS of the BS. In \cite{zhang2}, multiple users are involved and the practical case is considered, such as RIS with limited size can achieve a limited number of phase shifts, and a large-scale fading gain is needed on the BS. Based on the actual scenario, the authors propose the hybrid beamforming scheme in the RIS-based downlink system where a multi-antenna BS sends signals to users with the assumption that the BS-user link is unstable or even broken. Wu \emph{et al.} \cite{wu} studied the communication between a multi-antenna access point (AP) and multiple single-antenna users with RIS deployed in the single cell scenario, where both reflected AP-RIS-user link and direct AP-user link carried desired signals. Then, they jointly optimized active beamforming of AP and passive beamforming of RIS with discrete phase shifts due to hardware limitation. The authors aimed to solve the problem of minimizing transmission power at AP subjected to the individual SINR constraints. Also in the multiple-input single-output (MISO) network system, Huang \emph{et al.} \cite{huang} formulated an energy-efficient maximization problem, where power calculation was based on the phase quantization value and the number of reflection units. Downlink transmit powers and RIS continuous phase shifts were optimized under the constraints of minimum QoS requirements of users and maximum power.

For some other popular applications, RIS is also used to optimize performance. Wu \emph{et al.} \cite{wuwu} provided a basic introduction to RIS technology, detailed hardware architecture, competitive advantages compared to existed technologies, as well as a summary of its main applications in wireless communications. Yan \emph{et al.} \cite{yan} integrated passive beamforming and information transfer (PBIT) into wireless networks. It performed passive beamforming by adjusting the phase shifts of large intelligent surface (LIS) elements when they were turned on, which enhanced the communication quality between a single-antenna user and a M-antenna BS. Besides, spatial modulation was proposed to transmit private data of LIS. It should be emphasized that LIS has active components and passive components, and the passive operation is similar to RIS. In a RIS-aided unmanned aerial vehicle (UAV) communication system, Li \emph{et al.} \cite{li} jointly optimized UAV planned trajectory and RIS passive beamforming under the constraints of continuously controllable RIS phase shifts and practical UAV mobility. By designing RIS-spatial modulation and RIS-space shift keying schemes, Basar \emph{et al.} \cite{basar} applied RIS-based communication in the field of index modulation, and selected a specific receiving antenna index based on the information bits. Considering imperfect channel state information, Guo \emph{et al.} \cite{guo} investigated beamforming at the access point and the phase shifts of RIS elements with the goal of maximizing the weighted sum rate. There is also a case of high-frequency application of RIS. Chen \emph{et al.} \cite{chen} investigated the utilization of RIS in Terahertz communication for the indoor application scenario to make the sum rate significantly improved. However, RIS-assisted D2D communication has not been studied yet and it is very promising to carry out this research. Using the characteristic of passive beamforming, RIS will have a positive effect on the interference cancellation of D2D networks.

\section{System Model}\label{S3}

\subsection{System Description}\label{S3-1}

\begin{figure}[t]
\begin{center}
\includegraphics*[width=0.8\columnwidth,height=2.1in]{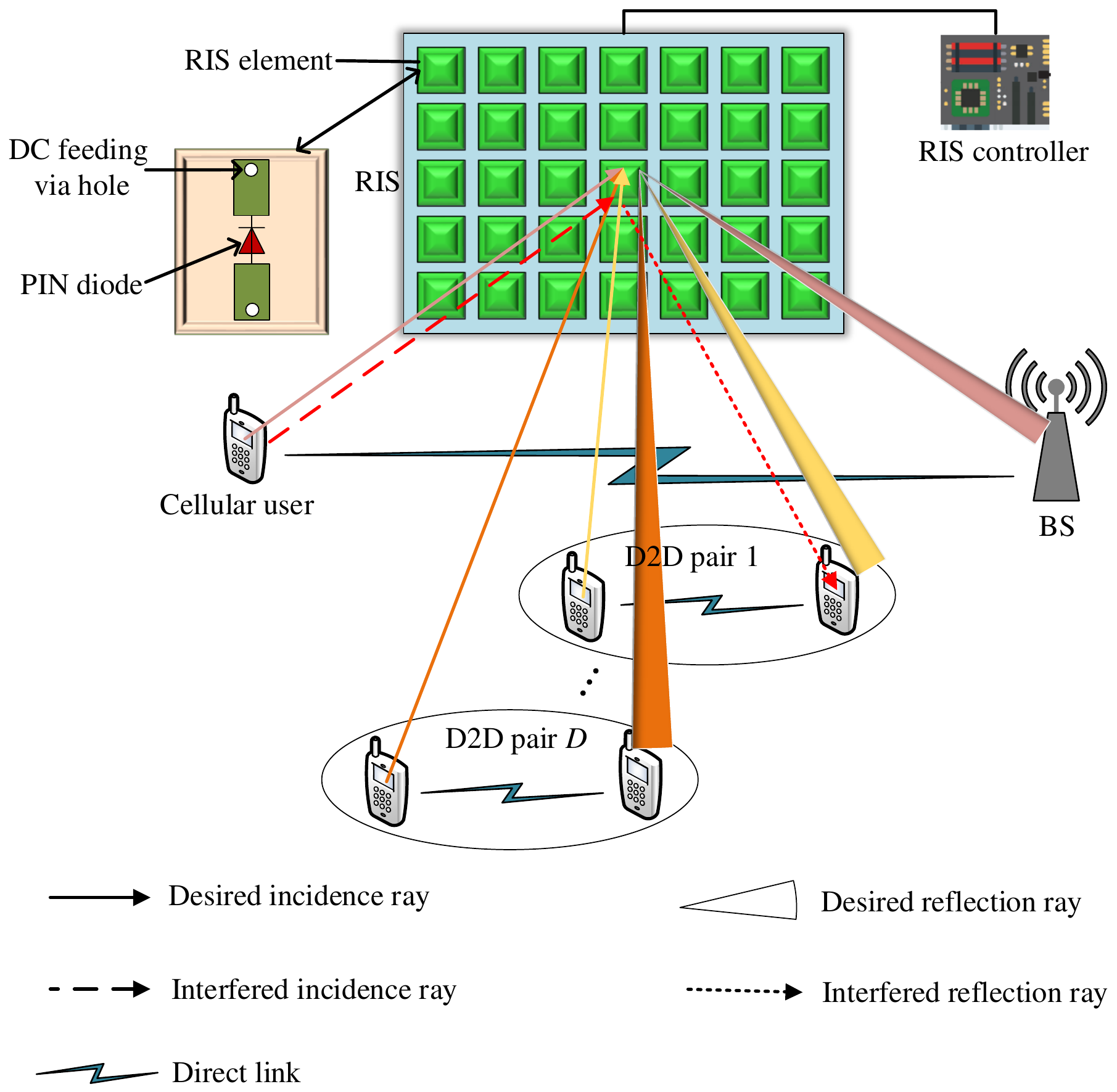}
\end{center}
\caption{System model for an uplink RIS-aided heterogeneous network.} \label{fig1}
\end{figure}
We consider an uplink single-cell RIS-aided heterogeneous network with one cellular user and multiple D2D pairs involved. As illustrated in Fig. \ref{fig1}, the RIS with extremely low cost is a passive reflecting device composed of a great quantity of built-in programmable elements. It reflects the signal from the transmitter and maps to the receiver with directional beam. Via the help of RIS, the virtual link between the source and destination is established.

The uplink communication between the BS and the cellular user is established with the assistance of RIS. At the same time, in the practical propagation environment, we assume the quality of this uplink channel is favorable and can support multiple D2D multiplexing. Thus, the direct transmission path of the cellular link is also included and is not easily deteriorated in this paper. Then, we also deploy $D$ D2D links, and they share the same frequency band with the cellular link. For D2D links, both the direct transmission mode and RIS-based reflecting mode are executed to enhance the received signal. Interference exists between all links, which becomes a bottleneck that limits system performance. In the system, there are a total of $D+1$ links and indexes are denoted as the set $\textbf{L}=\{1,2,...,D,D+1\}$.

In our investigated system, RIS is a uniform planar array consisting of $N\times N$ elements. When there is an incident wave, the phase shift of each element can be tuned in real time by voltage-controlled positive-intrinsic-negative (PIN) diodes with ON/OFF states. According to the regulation voltage, a certain phase shift will be provided by the metal plate. As we can seen from Fig. \ref{fig1}, the PIN diodes are embedded in a smart RIS controller. With reconfigurable characteristics, the propagation environment can be manipulated to alleviate interference between links and enhance the system performance. For simplicity, the range of phase shift of each element is constrained and we just take finite discrete values with equal quantization intervals between $[0, 2\pi]$. Assuming the number of quantization bits is $e$, there are $2^{e}$ patterns of phase shift values generated. The frequency response induced by the RIS element in the $l_{z}$-th row and the $l_{y}$-th column is $q_{l_{z},l_{y}}=e^{j\theta_{l_{z},l_{y}}}$ with phase shift $\theta_{l_{z},l_{y}}=\frac{2m_{l_{z},l_{y}}\pi}{2^{e}-1}, m_{l_{z},l_{y}}=\{0,1,...,2^{e}-1\}$, $1 \leq l_{z},l_{y} \leq N$, and $j$ denotes the imaginary unit. For each RIS element, the reflection operation is similar to multiply incident signal by the frequency response. Then, the reflector acts as a virtual source and forwards the composite signals to the desired receiver.

\subsection{Reflective and Direct Channel Model}\label{S3-2}
In this subsection, we first model the channel from the D2D transmitter to the receiver and that from the cellular user to the BS with the assistance of RIS. Since the distance between RIS elements is extremely small, the incoming signal does not randomly scatter into the open space after reaching the RIS, but is a superposition of spherical waves induced by many miniature scatters. At the same time, no additional signal forwarding and decoding process is required for reflected and refracted waves. In general, in view of the coupling effect between RIS elements, different from the traditional scattering-based two-hop relay forwarding channel propagation model, the signal does not propagate independently along two paths, from the transmitter to the RIS, and from the RIS to the receiver, but by passive reflection along a reflection path. Thus, we model the channel by using one-hop reflected ray. In addition, for access link, because the RIS reflected beam is directional, this transmission link can be regarded as a virtual line of sight (LoS) path, and the received signal is more favorable than other multipaths. For D2D links, reflection transmission is also virtual LoS path.

\begin{figure}[t]
\begin{center}
\includegraphics*[width=0.8\columnwidth,height=1.6in]{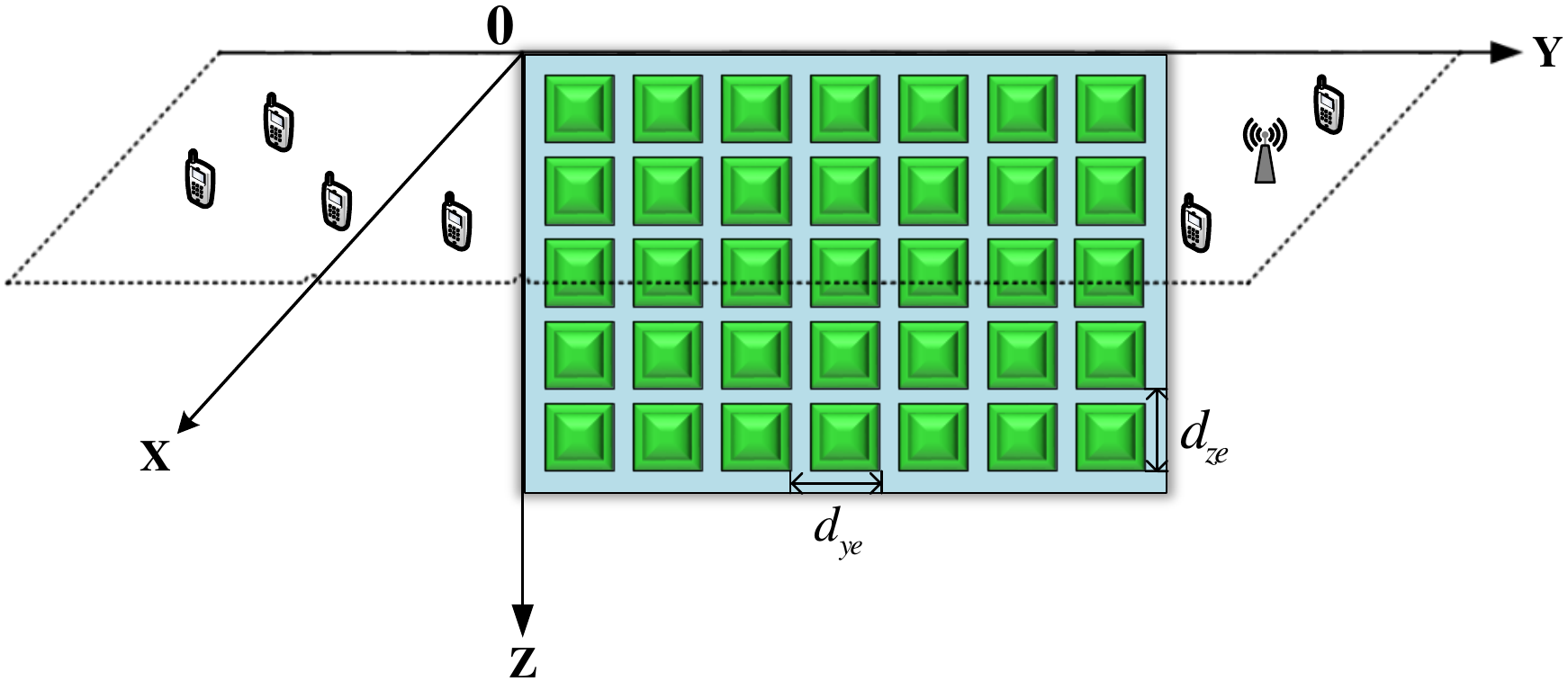}
\end{center}
\caption{Placement of RIS.} \label{fig2}
\end{figure}

As for the reflection-dominated channel, such as the user-RIS-BS link, we use the Rician distribution model, where this virtual LoS link is the LoS component, and the other paths are considered as non-LoS (NLoS) components. It is closely related to the position. Thus, we first establish three dimensional Cartesian coordinate system to represent the position of communication nodes. As shown in Fig. \ref{fig2}, RIS is placed on $Y$-$Z$ plane with the origin as the top left corner vertex, and the $Y$-axis and $Z$-axis as the alignment edges. $d_{ye}$ and $d_{ze}$ are the spacing between adjacent elements along $Y$ and $Z$ axes, respectively. It should be noted that, for each RIS element, we record its position with the vertex in the bottom right corner, i.e., $L_{\{l_{z},l_{y}\}}=(0,l_{y}d_{ye},l_{z}d_{ze})$. In addition, we assume that the BS, cellular user and D2D pairs are deployed on the $X$-$Y$ plane, which does not include the negative direction of the $X$ axis. This arrangement makes all communication nodes distributed on the front side of RIS. Take a cellular link or D2D link $i$ with transmitter $t_{i}$ and receiver $r_{i}$ as an example, the corresponding coordinates are denoted as $L_{t_{i}}=(t_{ix},t_{iy},0)$ and $L_{r_{i}}=(r_{ix},r_{iy},0)$. Based on given position, the distance from $t_{i}$ to the element $\{l_{z},l_{y}\}$, and from the element $\{l_{z},l_{y}\}$ to $r_{i}$, denoted as $DS^{l_{z},l_{y}}_{t_{i}}$ and $DS^{r_{i}}_{l_{z},l_{y}}$, respectively, can be easily obtained as follows,
\begin{equation}
DS^{l_{z},l_{y}}_{t_{i}}=\sqrt{(t_{ix})^{2}+(t_{iy}-l_{y}d_{ye})^{2}+(-l_{z}d_{ze})^{2}},
\label{eq1}
\end{equation}
\begin{equation}
DS^{r_{i}}_{l_{z},l_{y}}=\sqrt{(r_{ix})^{2}+(r_{iy}-l_{y}d_{ye})^{2}+(-l_{z}d_{ze})^{2}}.
\label{eq2}
\end{equation}

We assume that by communicating with the RIS controller, the BS can get the channel state information. Thus, the LoS component of reflection channel propagation from $t_{i}$ to $r_{i}$ associated with RIS element $\{l_{z},l_{y}\}$ can be expressed as
\begin{equation}
\tilde{h}^{r_{i},t_{i}}_{l_{z},l_{y}}=\sqrt{(DS^{l_{z},l_{y}}_{t_{i}}\cdot DS^{r_{i}}_{l_{z},l_{y}})^{-\alpha}}e^{-j
\frac{2\pi}{\lambda}(DS^{l_{z},l_{y}}_{t_{i}}+DS^{r_{i}}_{l_{z},l_{y}})},
\label{eq3}
\end{equation}
where $\lambda$ is the wavelength, and $\alpha$ is path-loss exponent. Adding NLoS components, the channel model is written as
\begin{equation}
\begin{aligned}
h^{r_{i},t_{i}}_{l_{z},l_{y}}&=\sqrt{\frac{\beta}{1+\beta}}\tilde{h}^{r_{i},t_{i}}_{l_{z},l_{y}}\\
&+\sqrt{\frac{1}{1+\beta}PL(DS^{l_{z},l_{y}}_{t_{i}}\cdot DS^{r_{i}}_{l_{z},l_{y}})}\tilde{h}^{r_{i},t_{i}}_{NLoS,{l_{z},l_{y}}},
\label{eq4}
\end{aligned}
\end{equation}
where $\beta$ is the Rician factor, $PL(\cdot)$ is the large-scale path loss function of NLoS transmission, and $\tilde{h}^{r_{i},t_{i}}_{NLoS,{l_{z},l_{y}}}\sim CN(0,1)$ is the small-scale fading.

Second, for the cellular link and the D2D links, we model the direct channel as $h_{r_{i},t_{i}}$, which is the baseband equivalent channel coefficient of the direct link. It behaves different with the reflected paths, and the channel model is
\begin{equation}
h_{r_{i},t_{i}}=h_{i}\sqrt{(DS^{r_{i}}_{t_{i}})^{-\alpha}},
\label{eq5}
\end{equation}
where $h_{i}$ is small-scale fading that obeys Nakagami-$m_{i}$ distribution with parameters $\{m_{i},\omega_{i}\}$, where $m_{i}$ is the fading depth parameter and $\omega_{i}$ is the average power in the fading signal. $(DS^{r_{i}}_{t_{i}})^{-\alpha}$ is large-scale path loss and the distance between $t_{i}$ and $r_{i}$ is also calculated based on the coordinates.

\subsection{Interference Analysis}\label{S3-3}

For receiver $r_{i}$ of link $i$, the intended signal received is the superposition of signal transmitted directly from transmitter $t_{i}$ and signal after traveling through the RIS reflection channel. Then, it treats all signals coming from remaining co-channel links as superposed interference. Summing all above interference together, we can express the received signal $s_{r_{i}}$ as
\begin{equation}
\begin{aligned}
s_{r_{i}}&=(h_{r_{i},t_{i}}+\sum\limits_{l_{z},l_{y}}h^{r_{i},t_{i}}_{l_{z},l_{y}}q_{l_{z},l_{y}})\sqrt{p_{i}}s_{t_{i}}\\
&+\sum\limits_{j\in \textbf{L}, j\neq i}(h_{r_{i},t_{j}}+\sum\limits_{l_{z},l_{y}}h^{r_{i},t_{j}}_{l_{z},l_{y}}q_{l_{z},l_{y}})\sqrt{p_{j}}s_{t_{j}}+w_{r_{i}},
\label{eq6}
\end{aligned}
\end{equation}
where $s_{t_{i}}$ is unit-power transmission information symbol corresponding to transmitter $t_{i}$, and $p_{i}$ is the transmission power of link $i$. To facilitate subsequent resolution, $\textbf{P}=[p_{1},p_{2},...,p_{D+1}]^{T}$ is a $(D+1)\times 1$ matrix representing the transmission power of all links $i=1,2,...,D+1$. $w_{r_{i}}$ is thermal Gaussian white noise with the components independently drawn from $CN(0,\sigma^{2})$. In order to simplify formula writing, we have some matrix definitions as given by
\begin{equation}
\textbf{F}=\sum\limits_{l_{z},l_{y}}q_{l_{z},l_{y}}\textbf{H}_{l_{z},l_{y}}.
\label{eq7}
\end{equation}
$\textbf{H}_{l_{z},l_{y}}$ is a $(D+1)\times (D+1)$ channel vector composed of reflection channel coefficients between all senders and receivers, $\{h^{r_{i},t_{j}}_{l_{z},l_{y}}, 1\leq i,j\leq D+1\}$. Similarly, $\textbf{H}^{L}$ is a matrix of direct channel coefficients with the same dimension of $(D+1) \times (D+1)$. It includes $h_{r_{i},t_{j}}, 1\leq i,j\leq D+1$. With the definition of these matrices, $s_{r_{i}}$ can be rewritten as
\begin{equation}
\begin{aligned}
s_{r_{i}}&=(\textbf{H}^{L}_{r_{i},t_{i}}+\textbf{F}_{r_{i},t_{i}})\sqrt{p_{i}}s_{t_{i}}\\
&+\sum\limits_{j\in \textbf{L}, j\neq i}(\textbf{H}^{L}_{r_{i},t_{j}}+\textbf{F}_{r_{i},t_{j}})\sqrt{p_{j}}s_{t_{j}}+w_{r_{i}}.
\label{eq8}
\end{aligned}
\end{equation}

Accordingly, the SINR received by links $i\in \textbf{L}$ is given as
\begin{equation}
\Gamma_{r_{i}}=\frac{|\textbf{H}^{L}_{r_{i},t_{i}}+\textbf{F}_{r_{i},t_{i}}|^2p_{i}}{\sum\limits_{j\in \textbf{L}, j\neq i}|\textbf{H}^{L}_{r_{i},t_{j}}+\textbf{F}_{r_{i},t_{j}}|^2p_{j}+\sigma^{2}}.
\label{eq9}
\end{equation}

According to the Shannon's capacity formula, the corresponding achievable transmission rates can be written as
\begin{equation}
R_{r_{i}}=\log_{2}\left(1+\frac{|\textbf{H}^{L}_{r_{i},t_{i}}+\textbf{F}_{r_{i},t_{i}}|^2p_{i}}{\sum\limits_{j\in \textbf{L}, j\neq i}|\textbf{H}^{L}_{r_{i},t_{j}}+\textbf{F}_{r_{i},t_{j}}|^2p_{j}+\sigma^{2}}\right).
\label{eq10}
\end{equation}

\section{Problem Formulation}\label{S4}
In this section, we first formulate our optimization problem based on the above system model, and then decompose the complex optimization problem in order to obtain a solution efficiently.

\subsection{Sum Rate Maximization Problem Formulation}\label{S4-1}
The optimization objective is maximizing the sum rate, where the optimization variables are the phase shift values of all the RIS elements and the transmission power of all the links. Moreover, we record the phase shift values applied to all elements as a set $\bm{\Theta}=\{\theta_{l_{z},l_{y}}, 1 \leq l_{z},l_{y} \leq N\}$, and the optimization problem can be written as follows.
\begin{equation}\label{eq11}
\begin{split}
&\max\limits_{\textbf{P},\bm{\Theta}}\ \sum\limits_{i=1}^{D+1}R_{r_{i}}   \\
&s.t. \\
& (a)\ \Gamma_{r_{i}}\geq \gamma_{min},  \  \forall  i=1,2,...,D+1, \\
& (b)\ 0\leq p_{i}\leq P_{max},  \  \forall  i=1,2,...,D+1,\\
& (c)\ q_{l_{z},l_{y}}=e^{j\theta_{l_{z},l_{y}}}, \ \theta_{l_{z},l_{y}}=\frac{2m_{l_{z},l_{y}}\pi}{2^{e}-1}, \\
& \ \ \ \  m_{l_{z},l_{y}}=\{0,1,...,2^{e}-1\}, 1 \leq l_{z},l_{y} \leq N,
\end{split}
\end{equation}
where constraint (a) indicates the minimum SINR requirements for cellular link and D2D links to ensure the QoS. Then, we limit the transmission power as in constraint (b) to effectively manage the interference. In constraint (c), the amplitude reflection coefficient of each element is 1, and the phase shift is a discrete variable. The above formulated problem is a mixed integer non-convex non-linear optimization problem. Both the objective function and constraint (a) involve the non-convex SINR formula, in which two variables, $\textbf{P}$ and $\bm{\Theta}$, are coupled. However, this coupling relationship is hard to eliminate, which makes the problem challenging and difficult to solve.

\subsection{Problem Decomposition}\label{S4-2}
To solve the optimization problem in (\ref{eq11}) efficiently, we decouple it into two easy-to-solve sub-problems.

\emph{1) Power Allocation}: This sub-problem is to allocate the appropriate transmission power within the power constraint to maximize the objective function on the basis that the SINR constraints are satisfied. When the other variable $\bm{\Theta}$ is fixed, the problem in (\ref{eq11}) about transmission power $\textbf{P}$ can be written as
\begin{equation}\label{eq12}
\begin{split}
&\max\limits_{\textbf{P}}\ \sum\limits_{i=1}^{D+1}R_{r_{i}}   \\
&s.t. \\
& (a)\ \Gamma_{r_{i}}\geq \gamma_{min},  \  \forall  i=1,2,...,D+1, \\
& (b)\ 0\leq p_{i}\leq P_{max},  \  \forall  i=1,2,...,D+1.
\end{split}
\end{equation}

\emph{2) Discrete Phase Shift Optimization}: We fix $\textbf{P}$ and this sub-problem is to provide an efficient phase shift from $2^e$ values for each RIS element. After that, the optimization problem is transformed to
\begin{equation}\label{eq13}
\begin{split}
&\max\limits_{\bm{\Theta}}\ \sum\limits_{i=1}^{D+1}R_{r_{i}}   \\
&s.t. \\
& (a)\ \Gamma_{r_{i}}\geq \gamma_{min},  \  \forall  i=1,2,...,D+1, \\
& (b)\ q_{l_{z},l_{y}}=e^{j\theta_{l_{z},l_{y}}}, \ \theta_{l_{z},l_{y}}=\frac{2m_{l_{z},l_{y}}\pi}{2^{e}-1}, \\
&\ \ \ \ \ m_{l_{z},l_{y}}=\{0,1,...,2^{e}-1\}, 1 \leq l_{z},l_{y} \leq N.
\end{split}
\end{equation}

In view of these two sub-problems, we will design efficient algorithms to solve them in the next section, and then achieve the goal of solving the optimization problem.

\section{Sum Rate Maximization}\label{S5}
In this section, we first solve the aforementioned two sub-problems, and then we propose a sum rate maximization algorithm, which adopts the alternating method to solve these two sub-problems iteratively until the algorithm converges and a sub-optimal solution is obtained.

\subsection{Power Allocation Sub-problem Algorithm Design}\label{S5-1}

Although the integer discrete variables are removed, the sub-problem is still non-convex and non-linear with respect to $\textbf{P}$. Motivated by the special structure, we use some simple mathematical transformations. In the objective function, both the cellular transmission rate and D2D transmission rate can be written as the difference of two concave functions, and the objective function of this sub-problem is equivalent to
\begin{equation}
\begin{aligned}
&\max\limits_{\textbf{P}}\sum\limits_{i=1}^{D+1}R_{r_{i}}\\
=&\max\limits_{\textbf{P}}\sum\limits_{i=1}^{D+1}\log_{2}\left(1+\frac{|\textbf{H}^{L}_{r_{i},t_{i}}+\textbf{F}_{r_{i},t_{i}}|^{2}p_{i}}{\sum\limits_{j\in \textbf{L}, j\neq i}|\textbf{H}^{L}_{r_{i},t_{j}}+\textbf{F}_{r_{i},t_{j}}|^{2}p_{j}+\sigma^{2}}\right)\\
=&-\min\limits_{\textbf{P}}\sum\limits_{i=1}^{D+1}\Bigg[\lg\Bigg(\sum\limits_{j\in \textbf{L}, j\neq i}|\textbf{H}^{L}_{r_{i},t_{j}}+\textbf{F}_{r_{i},t_{j}}|^{2}p_{j}+\sigma^{2}\Bigg)\\
&-\lg\Bigg(|\textbf{H}^{L}_{r_{i},t_{i}}\!\!+\!\textbf{F}_{r_{i},t_{i}}|^{2}p_{i}\!+\!\!\!\!\sum\limits_{j\in \textbf{L}, j\neq i}|\textbf{H}^{L}_{r_{i},t_{j}}\!\!+\!\textbf{F}_{r_{i},t_{j}}|^{2}p_{j}\!+\!\sigma^{2}\Bigg)\Bigg].
\label{eq14}
\end{aligned}
\end{equation}

Based on the above mathematical decomposition, for a link $i$, we regard these two logarithmic function terms in the new objective function as two separated functions, denoted as $g_{i}(\textbf{P})$ and $\varphi_{i}(\textbf{P})$, which is known as the difference of two concave functions.
\begin{equation}
g_{i}(\textbf{P})=\lg\left(\sum\limits_{j\in \textbf{L}, j\neq i}|\textbf{H}^{L}_{r_{i},t_{j}}+\textbf{F}_{r_{i},t_{j}}|^{2}p_{j}+\sigma^{2}\right).
\label{eq15}
\end{equation}
\begin{equation}
\begin{aligned}
\varphi_{i}(\textbf{P})&=\lg\Bigg(|\textbf{H}^{L}_{r_{i},t_{i}}+\textbf{F}_{r_{i},t_{i}}|^{2}p_{i}\\
&+\sum\limits_{j\in \textbf{L}, j\neq i}|\textbf{H}^{L}_{r_{i},t_{j}}+\textbf{F}_{r_{i},t_{j}}|^{2}p_{j}+\sigma^{2}\Bigg).
\label{eq16}
\end{aligned}
\end{equation}

Then, problem (\ref{eq12}) is a difference of concave functions problem, and is equivalently written as
\begin{equation}\label{eq17}
\begin{split}
&\min\limits_{\textbf{P}}\ \sum\limits_{i=1}^{D+1}f_{i}(\textbf{P})\triangleq g_{i}(\textbf{P})-\varphi_{i}(\textbf{P})   \\
&s.t. \\
& (a)\ g_{i}(\textbf{P})-\varphi_{i}(\textbf{P}) \leq -\lg(\gamma_{min}+1),  \  \forall  i=1,2,...,D+1, \\
& (b)\ 0\leq p_{i}\leq P_{max},  \  \forall  i=1,2,...,D+1.
\end{split}
\end{equation}
So far, this DC problem is still not easy to solve. Thus, we use the first-order Taylor expansion to approximate it as a convex function, i.e., to provide an upper bound for the objective function that needs to be minimized, and then gradually approach the optimal solution from the upper bound. This transformation has been mathematically proven to achieve faster convergence rate. Finally, both the objective function and the constraints are convex with optimization variable $\textbf{P}$. The Taylor expansion of $g_{i}(\textbf{P})$ in the $n$-th iteration is written as follows.
\begin{equation}
g_{i}(\textbf{P})=g_{i}(\textbf{P}^{(n)})+\sum\limits_{k=1}^{D+1}\frac{\partial g_{i}(\textbf{P})}{\partial p_{k}}\Big|_{\textbf{P}=\textbf{P}^{(n)}}(p_{k}-p^{(n)}_{k}).
\label{eq18}
\end{equation}

Substituting (\ref{eq18}) into problem (\ref{eq17}) and let
\begin{equation}
f^{(n)}_{i}(\textbf{P})=g_{i}(\textbf{P}^{(n)})+\sum\limits_{k=1}^{D+1}\frac{\partial g_{i}(\textbf{P})}{\partial p_{k}}\Big|_{\textbf{P}=\textbf{P}^{(n)}}(p_{k}-p^{(n)}_{k})-\varphi_{i}(\textbf{P}),
\label{eq19}
\end{equation}
the problem can be further simplified into
\begin{equation}
\begin{split}
\begin{aligned}
&\min\limits_{\textbf{P}}\ \sum\limits_{i=1}^{D+1}f^{(n)}_{i}(\textbf{P})  \\
&s.t. \\
& (a)\ f^{(n)}_{i}(\textbf{P}) \leq -\lg(\gamma_{min}+1), \ \forall  i=1,2,...,D+1, \\
& (b)\ 0\leq p_{i}\leq P_{max},  \  \forall  i=1,2,...,D+1.
\end{aligned}
\end{split}
\label{eq20}
\end{equation}

Therefore, this problem is converted into a convex problem, which can be solved by the Lagrangian method. Specifically, we first construct a Lagrangian unconstrained function for the problem. Then, by utilizing an iterative gradient descent method, the objective function is monotonically decreasing, so that it can converge to a static point. At the $n$-th iteration, the Lagrange function on optimization variable $\textbf{P}$ is yielded as follows,
\begin{equation}
L^{(n)}(\textbf{P},\bm{\lambda}^{(n)})
=\sum\limits_{i=1}^{D+1}f^{(n)}_{i}(\textbf{P})+\lambda^{(n)}_{i}[f^{(n)}_{i}(\textbf{P})+\lg(\gamma_{min}+1)],
\label{eq21}
\end{equation}
where $\lambda^{(n)}_{i}, i=1,2,...,D+1$ refer to the Lagrangian multipliers corresponding to constraint (a) in problem (\ref{eq20}) at the $n$-th iteration. Meanwhile, by solving the Lagrange function in (\ref{eq21}), we obtain transmission power $\textbf{P}^{(n+1)}$ and substitute it into (\ref{eq21}). Then, the Lagrangian function for Lagrangian multiplier $\bm{\lambda}$ is given as,
\begin{equation}
\begin{aligned}
&L^{(n)}(\textbf{P}^{(n+1)},\bm{\lambda})\\
&=\sum\limits_{i=1}^{D+1}\Bigg[g_{i}(\textbf{P}^{(n)})+\sum\limits_{k=1}^{D+1}\frac{\partial g_{i}(\textbf{P})}{\partial p_{k}}\Big|_{\textbf{P}=\textbf{P}^{(n)}}(p^{(n+1)}_{k}-p^{(n)}_{k})\\
&-\varphi_{i}(\textbf{P}^{(n+1)})\Bigg]\\
&+\sum\limits_{i=1}^{D+1}\lambda_{i}\Bigg[g_{i}(\textbf{P}^{(n)})+\sum\limits_{k=1}^{D+1}\frac{\partial g_{i}(\textbf{P})}{\partial p_{k}}\Big|_{\textbf{P}=\textbf{P}^{(n)}}(p^{(n+1)}_{k}-p^{(n)}_{k})\\
&-\varphi_{i}(\textbf{P}^{(n+1)})+\lg(\gamma_{min}+1)\Bigg].
\label{eq22}
\end{aligned}
\end{equation}

\begin{algorithm}[t]
\caption{DC Programming for Power Allocation}
\label{alg1}
\begin{algorithmic}[1]
\REQUIRE
  $n=0, \delta^{(0)}=50, \mu^{(0)}=100, \lambda^{(0)}_{i}=100, \forall i=1,2,...,D+1$
\ENSURE
  $\textbf{P}^{\ast}$
\STATE Construct an auxiliary optimization problem (\ref{eq20}) using DC programming theory and Taylor expansion of multivariate function;
\STATE Define Lagrangian unconstrained function $L^{(n)}(\textbf{P},\bm{\lambda}^{(n)})$ as (\ref{eq21});
\STATE Solve the optimization problem about $p_{i}, i=1,2,...,D+1$;
\STATE $p^{(n+1)}_{i}=\left(p^{(n)}_{i}-\delta^{(n)}\frac{\partial L^{(n)}(\textbf{P},\bm{\lambda}^{(n)})}{\partial p_{i}}\big|_{p_{i}=p^{(n)}_{i}}\right)^{+}$;
\IF {$p^{(n+1)}_{i}> P_{max}$}
\STATE $p^{(n+1)}_{i}= P_{max}$;
\ENDIF
\IF {$\delta^{(n)}>1$}
\STATE $\delta^{(n+1)}=\frac{\delta^{(n)}}{2}$;
\ENDIF
\STATE $\lambda^{(n+1)}_{i}=\left(\lambda^{(n)}_{i}-\mu^{(n)}\left(\frac{\partial L^{(n)}(\textbf{P}^{(n+1)},\bm{\lambda})}{\partial \lambda_{i}}\big|_{\lambda_{i}=\lambda^{(n)}_{i}}\right)^{+}\right)^{+}, i=1,2,...,D+1$;
\IF {$\mu^{(n)}>1$}
\STATE $\mu^{(n+1)}=\frac{\mu^{(n)}}{2}$;
\ENDIF
\IF {$|R(\textbf{P}^{(n+1)})-R(\textbf{P}^{(n)})|<\epsilon$}
\STATE $\textbf{P}^{\ast}=\textbf{P}^{(n+1)}$;
\ELSE
\STATE $n=n+1$, and go to line 1;
\ENDIF
\end{algorithmic}
\end{algorithm}

The details of DC programming for power allocation are shown in Algorithm \ref{alg1}.
We initialize Lagrangian multipliers $\lambda_{i}, \forall i=1,2,...,D+1$. Besides, the initial values of corresponding gradient descent step sizes $\delta^{(0)}$ and $\mu^{(0)}$ for transmission power and $\lambda_{i}$ are also given, respectively. Then, we construct an optimization problem in (\ref{eq20}) using the DC programming and multivariate Taylor expansion method introduced before in order to approximate the non-convex original problem. In step 2, the Lagrangian unconstrained function is defined for the transformed convex problem with $\bm{\lambda}^{(n)}$ fixed. As for optimization variable $p_{i}$, the gradient descent method computes the optimal gradient descent direction as the opposite direction of the first-order partial derivative, and the suitable step size $\delta$ is selected to gradually decrease with the number of iterations based on the given initial value. Combined with the basic constraints of transmission power, $0\leq p_{i}\leq P_{max}, \forall  i=1,2,...,D+1$, the optimal values for this iteration or required initial values for the next iteration are generated. From step 11, we turn to optimize $\lambda_{i}$. It should be noted that optimizing $\lambda_{i}$ requires substituting the optimal value of $p_{i}$ solved from steps 4-7. Until in a certain iteration, the difference between two sum rate results is less than a preset threshold. Then, we output the optimal solution $\textbf{P}^{\ast}$. Otherwise, we proceed to the next iteration with the power $\textbf{P}^{(n+1)}$ obtained.
\begin{algorithm}[t]
\caption{Local Search for Phase Shift}
\label{alg2}
\begin{algorithmic}[1]
\REQUIRE
  the number of quantization bits $e$
\ENSURE
  $\bm{\Theta}^{\ast}$
\FOR {$l_{z}=1:N$}
\FOR {$l_{y}=1:N$}
\STATE Assign all possible values to $\theta_{l_{z},l_{y}}$, and select the value maximizing the sum rate on the premise that constraint $(\ref{eq13}a)$ is satisfied, denoted as $\theta^{\ast}_{l_{z},l_{y}}$;
\STATE $\theta_{l_{z},l_{y}}=\theta^{\ast}_{l_{z},l_{y}}$;
\ENDFOR
\ENDFOR
\end{algorithmic}
\end{algorithm}

\begin{algorithm}[t]
\caption{Sum Rate Maximization}
\label{alg3}
\begin{algorithmic}[1]
\REQUIRE
  $\epsilon=10^{-3}, \varrho=0, p^{(0)}_{i}=P_{max}, \forall i=1,2,...,D+1$, the number of quantization bits $e$, randomly generate $\bm{\Theta}, \bm{\Theta}^{\ast}=\bm{\Theta}$
\STATE Given $\bm{\Theta}^{\ast}$ update $\textbf{P}$ using Algorithm \ref{alg1};
\STATE Given $\textbf{P}^{\ast}$ update $\bm{\Theta}$ using Algorithm \ref{alg2};
\IF {$|R^{(\varrho+1)}-R^{(\varrho)}|<\epsilon$}
\STATE $R^{\ast}=R^{(\varrho+1)}$;
\STATE \textbf{Output} $\textbf{P}^{\ast},\bm{\Theta}^{\ast},R^{\ast}$;
\ELSE
\STATE $\varrho=\varrho+1$, and go to line 1;
\ENDIF
\end{algorithmic}
\end{algorithm}

\subsection{Discrete Phase Shift Optimization Sub-problem Algorithm Design}\label{S5-2}
When fixing the transmission power $\textbf{P}$, only the simple power range constraint has been removed. The objective function and constraints about $\bm{\Theta}$ are still non-convex. Besides, $\bm{\Theta}$ contains a series of discrete variables, and the range available for each phase shift depends on the RIS quantization bits. Considering the complexity, we use the local search method as shown in Algorithm \ref{alg2} to solve this problem. Specifically, keeping the other $N^{2}-1$ phase shift values fixed, for each element $\theta_{l_{z},l_{y}}$, we traverse all possible values and choose the optimal one without violating the SINR constraints. Then, use this optimal solution $\theta^{\ast}_{l_{z},l_{y}}$ as the new value of $\theta_{l_{z},l_{y}}$ for the optimization of another phase shift, until all phase shifts in the set $\bm{\Theta}$ are fully optimized.

\subsection{Sum Rate Maximization}\label{S5-3}
We summarize the above power allocation sub-problem algorithm and discrete phase shift optimization sub-problem algorithm, and propose the sum rate maximization algorithm. As shown in Algorithm \ref{alg3}, we set all links to transmit at the maximum power budget in the initial state, and randomly generate a phase shift matrix. Then, we update the transmission power and phase shift in an alternating manner until the algorithm converges, i.e., the system transmission rate difference between two iterations is less than a certain threshold, $|R^{(\varrho+1)}-R^{(\varrho)}|<\epsilon$.

\subsection{Convergence and Complexity Analysis}\label{S5-4}
\emph{1) Convergence}:
In the sum rate maximization algorithm, iterative calculations of two sub-problems are involved. When $\bm{\Theta}$ is fixed, $\textbf{P}$ is solved using Algorithm \ref{alg1}. For Algorithm \ref{alg1}, first, for the objective function $f_{i}(\textbf{P})$, we construct an auxiliary function using DC programming theory and multivariate Taylor expansion at the $n$-th iteration as $f^{(n)}_{i}(\textbf{P})$. Then, $f_{i}(\textbf{P}^{(n)})=f^{(n)}_{i}(\textbf{P}^{(n)})$ holds. In each iteration, we can get the optimal solution $\textbf{P}^{(n+1)}$ of optimization problem (20) using gradient descent method. Thus, $f^{(n)}_{i}(\textbf{P}^{(n)})\geq f^{(n)}_{i}(\textbf{P}^{(n+1)})$, that is, the system transmission rate has increased. Since $\lg\left(\sum\limits_{j\in \textbf{L}, j\neq i}|\textbf{H}^{L}_{r_{i},t_{j}}+\textbf{F}_{r_{i},t_{j}}|^{2}p_{j}+\sigma^{2}\right)$ is concave,
\begin{equation}
\begin{aligned}
&\lg\left(\sum\limits_{j\in \textbf{L}, j\neq i}|\textbf{H}^{L}_{r_{i},t_{j}}+\textbf{F}_{r_{i},t_{j}}|^{2}p_{j}+\sigma^{2}\right) \\
& \leq g_{i}(\textbf{P}^{(n)})+\sum\limits_{k=1}^{D+1}\frac{\partial g_{i}(\textbf{P})}{\partial p_{k}}\Big|_{\textbf{P}=\textbf{P}^{(n)}}(p_{k}-p^{(n)}_{k}),
\label{eq23}
\end{aligned}
\end{equation}
and the inequality
\begin{equation}
f^{(n)}_{i}(\textbf{P}) \geq f_{i}(\textbf{P}),
\label{eq24}
\end{equation}
is also satisfied. Based on the above analysis, we can conclude
\begin{equation}
f_{i}(\textbf{P}^{(n)})=f^{(n)}_{i}(\textbf{P}^{(n)})\geq f^{(n)}_{i}(\textbf{P}^{(n+1)})\geq f_{i}(\textbf{P}^{(n+1)}).
\label{eq25}
\end{equation}
Thus, as the number of iterations increases, the original objective function $f_{i}(\textbf{P})$ also decreases monotonically, and the convergence of the original problem is proved. Next, we need to show in the following theorem that the solution obtained by the approximated problem is also a feasible solution to the original one.
\begin{theorem}
$g_{i}(\textbf{P})$ and $\varphi_{i}(\textbf{P})$ are continuously differentiable in the domain of transmission power. Then, solving the auxiliary problem can get a stationary point of the original problem.
\label{the1}
\end{theorem}
The first thing to explain is that in any problem domain, the stationary point of a function $f_{i}(\textbf{P})$ is a point $\textbf{\={P}}$ that satisfies the following conditions,
$\frac{\partial f_{i}(\textbf{P})}{\partial p_{k}}\big|_{\textbf{P}=\textbf{\={P}}}(p_{k}- \bar{p}_{k})\geq 0, \forall k=1,2,...,D+1$. Since the $f_{i}(\textbf{P})$ can be converged, when the number of iterations approaches infinity, $\textbf{P}^{(n)}=\textbf{P}^{(n+1)}$, and both of them are optimal solutions. Furthermore, $f_{i}(\textbf{P}^{(n)})=f^{(n)}_{i}(\textbf{P}^{(n)})\geq f^{(n)}_{i}(\textbf{P}^{(n+1)})\geq f_{i}(\textbf{P}^{(n+1)})$ are changed to $f_{i}(\textbf{P}^{(n)})=f^{(n)}_{i}(\textbf{P}^{(n)})= f^{(n)}_{i}(\textbf{P}^{(n+1)})= f_{i}(\textbf{P}^{(n+1)})$. $\textbf{P}^{(n)}$ and $\textbf{P}^{(n+1)}$ are optimal solutions of $f^{(n)}_{i}(\textbf{P})$. Let $\textbf{P}^{(n)}$ be the final solution, $f^{(n)}_{i}(\textbf{P})$ is convex, and $\textbf{P}^{(n)}$ is the stationary point of $f^{(n)}_{i}(\textbf{P})$. Thus, $\frac{\partial f^{(n)}_{i}(\textbf{P})}{\partial p_{k}}\big|_{\textbf{P}=\textbf{P}^{(n)}}(p_{k}-p^{(n)}_{k})\geq 0$, that is, $\frac{\partial f_{i}(\textbf{P})}{\partial p_{k}}\big|_{\textbf{P}=\textbf{P}^{(n)}}(p_{k}-p^{(n)}_{k})\geq 0$. Thus, $\textbf{P}^{(n)}$ is also the stationary point of a function $f_{i}(\textbf{P})$. Finally, we have completed the proof.

After proving that the auxiliary problem is feasible, we switch back to Algorithm \ref{alg1}. The goal is to maximize system transmission rate with solving the Lagrangian unconstrained function obtained by mathematical transformation. In line 1 of Algorithm \ref{alg3}, with $\bm{\Theta}^{\ast}$ given at the $\varrho$-th iteration, $R({\textbf{P}^{\ast}}^{(\varrho+1)},\bm{\Theta}^{\ast})\geq R({\textbf{P}^{\ast}}^{(\varrho)},\bm{\Theta}^{\ast})$ is satisfied. Then, we turn to the second step of Algorithm \ref{alg3}. When $\textbf{P}^{\ast}$ is fixed, $\bm{\Theta}$ is updated using Algorithm \ref{alg2}. The local search algorithm aims at maximizing the sum rate, and will eventually get ${\bm{\Theta}^{\ast}}^{(\varrho+1)}$ corresponding to fixed $\textbf{P}^{\ast}$. In other words, the optimal phase shifts ${\bm{\Theta}^{\ast}}^{(\varrho+1)}$ must be greater than or equal to the initial value of $\bm{\Theta}$, which is also the value of $\bm{\Theta}^{\ast}$ obtained from the $\varrho$-th iteration of Algorithm \ref{alg3}. Thus, $R(\textbf{P}^{\ast},{\bm{\Theta}^{\ast}}^{(\varrho+1)})\geq R(\textbf{P}^{\ast},{\bm{\Theta}^{\ast}}^{(\varrho)})$. In summary, $R({\textbf{P}^{\ast}}^{(\varrho+1)},{\bm{\Theta}^{\ast}}^{(\varrho+1)})\geq R({\textbf{P}^{\ast}}^{(\varrho)},{\bm{\Theta}^{\ast}}^{(\varrho)})$. We can see the objective function of original optimization problem is non-decreasing. In addition, the number of discrete phase shifts is limited, and the continuous transmission power is also constrained by the upper and lower bounds, which make the problem of maximizing the sum rate bounded and the output solutions guaranteed. Therefore, we have completed the proof of the convergence of the sum rate maximization algorithm.

\emph{2) Complexity}:
The complexity of sum rate maximization algorithm is not only related to the number of iterations, which has been set to $N_{outer}$ to achieve convergence condition $|R^{(\varrho+1)}-R^{(\varrho)}|<\epsilon$, but also related to the complexity of power allocation sub-problem and phase shifts sub-problem. For the former one, the number of gradient updates and the optimization of transmission power of all links in each gradient iteration create complexity. The number of cellular link and D2D links is $D+1$, and we denote the number of iterations of gradient descent method as $N_{inner}$. Consequently, the complexity of power allocation sub-problem is $O(N_{inner}*(D+1))$. For the latter one, for each element $\{l_{z},l_{y}\}$, the local search algorithm keeps the remaining phase shifts unchanged, selects the best one among the $2^{e}$ phase shifts and assigns value for $\theta_{l_{z},l_{y}}$. Combining RIS is a $N\times N$ planar array, the complexity of this part is $O(N^{2}*2^{e})$. Therefore, we get the complexity of the proposed sum rate maximization algorithm as $O(N_{outer}*(N_{inner}*(D+1)+N^{2}*2^{e}))$.

\section{Performance Evaluation}\label{S6}
In this section, by comparison with other benchmark algorithms, we obtain numerical results under various representative parameters to verify the effectiveness of our proposed scheme, and investigate the impact of different parameters on system performance.

\subsection{Simulation Setup}\label{S6-1}
In the simulation, we first introduce the network topology. All links are deployed in a rectangular area with four points $(0,-100,0), (0,100,0), (100,100,0)$ and $(100,-100,0)$ as vertices in a three-dimensional coordinate system. The cellular user and BS are distributed on both sides of the positive semi-axis of $X$ with a distance of 10. Then, RIS lies on the $Y$-$Z$ plane, and the size is very small. We assign a value of 0.03 to both RIS element separations $d_{ye}$ and $d_{ze}$. It is equivalent to that cellular user and BS are distributed on both sides of RIS. The direct distance of the cellular link is controlled so that it has the ability to reuse D2D links. All D2D users, regardless of the transmitter or receiver, are uniformly and randomly scattered in the rectangular area. In accordance with the realistic case, we set the maximum distance between two D2D devices to $10$. Furthermore, we give some parameter settings. For large-scale fading, we adopt the path loss model in the urban micro (UMi) scenario \cite{pathloss}. $PL(DS)=22.0*\lg(DS)+28.0+20*\lg(f_{c})$ in the LoS case, and $PL(DS)=36.7*\lg(DS)+22.7+26*\lg(f_{c})$ in the NloS case, where $f_{c}$ is the centre frequency in GHz and $DS$ is distance in meter. In this paper, we use the millimeter-wave (mm-wave) band with a center frequency of 28 GHz for simulation. The mm-wave noise power spectral density is -134 dBm/MHz. Besides, the Rician factor of the reflection channel is set to be 4, and the channel parameters of Nakagami distribution $m_{s}=3, \omega_{s}=1/3$. Without loss of generality, the maximum transmission power $P_{max}$ is 23 dBm, and all individuals have the same minimum SINR $\gamma_{min}=5$ dB. Finally, not only for the power allocation algorithm, but also for the sum rate maximization algorithm, we set the termination iteration threshold as $10^{-3}$. The basic system parameters are set as above, unless specified later.

To show the system performance of the \textbf{Proposed-algorithm}, we compare it with the following algorithms.
\begin{itemize}
\item \textbf{Maximum Power Transmission (MPT)}: for this algorithm, all cellular and D2D links are operated at the maximum transmission power without any power allocation. The purpose of optimizing system performance and mitigating interference as much as possible is to adjust the phase shifts of RIS elements by applying local search algorithm. Compared with the proposed scheme, the MRT omits the design of transmission power. However, transmitting at maximum power makes it comparable with the proposed algorithm.
\item \textbf{Random Phase Shift (RPS)}: this algorithm randomly selects a feasible phase shift in the initial state, and RIS is configured. After that, no reconfiguration is performed. Then, the maximum power is used as the initial transmission power for all links. By using the power allocation sub-problem algorithm proposed in this paper, the goal of maximizing sum rate is approached. This scheme is relatively poor. The reason behind this is it deals with only one optimization variable. Meanwhile, the value of another optimization variable, phase shift, has no advantage.
\item \textbf{Without-RIS}: this scheme does not use RIS for signal reflection, and only direct propagation exists. Thus, it just involves power optimization.
\end{itemize}

\subsection{Comparison with Other Schemes}\label{S6-1}
\begin{figure}[t]
\begin{center}
\includegraphics*[width=0.8\columnwidth,height=2.2in]{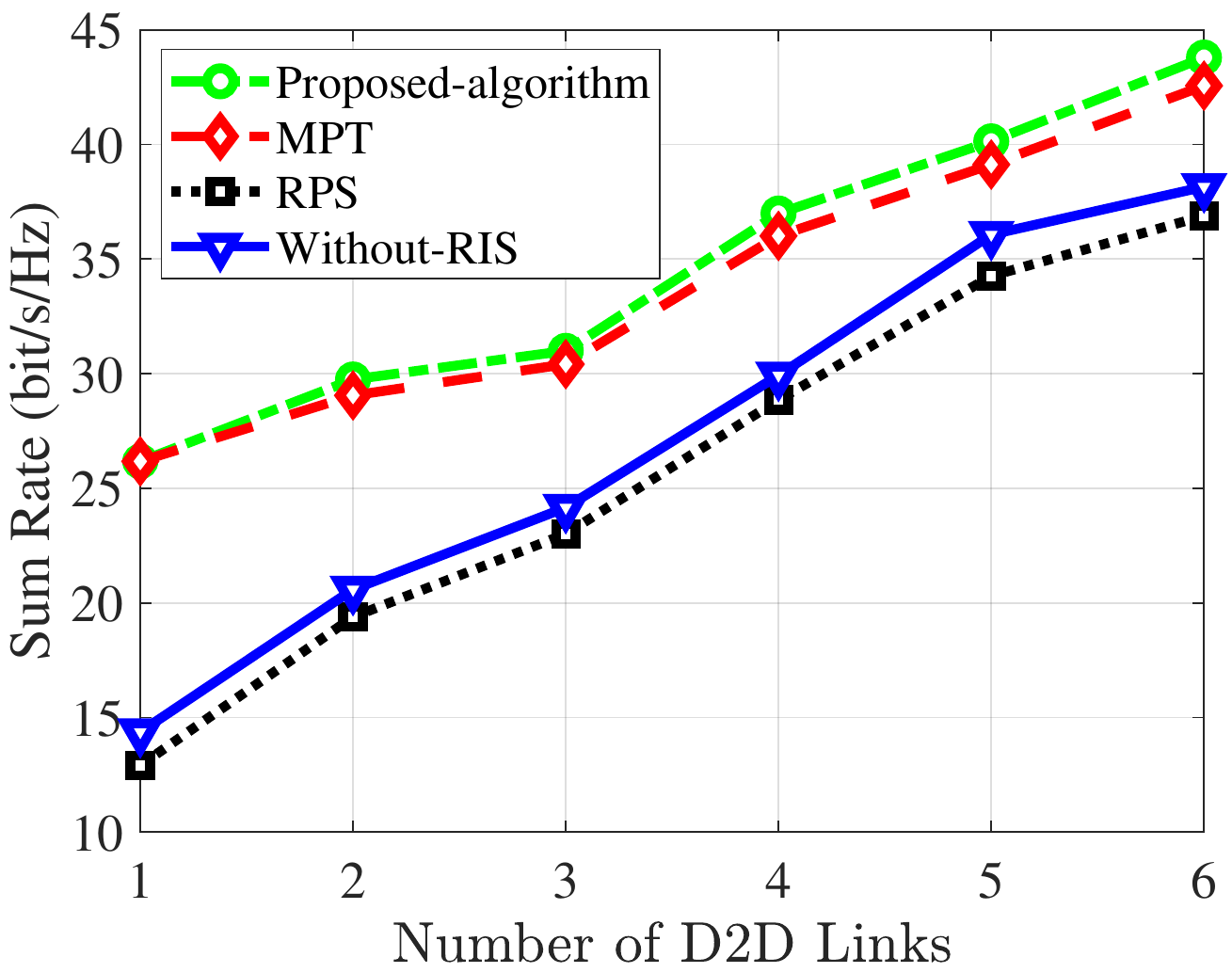}
\end{center}
\caption{Sum rate versus the number of D2D links.} \label{fig3}
\end{figure}

In Fig. \ref{fig3}, we set $N=4, e=3$, and provide the sum rate performance curves of four schemes as the number of D2D links changes from 1 to 6. It is observed that the proposed algorithm achieves the highest transmission capacity, and the sum rates of all schemes monotonically increase with the number of D2D links. As more D2D links are deployed, the gap between the proposed scheme and the MPT is increasing, which shows that the effect of power control is gradually obvious. Then, a noticeable difference is observed between the proposed one and the other two schemes, Without-RIS and RPS. This can be attributed to the adoption of RIS. Not using RIS to assist communications or using RIS but not performing phase shift optimization will suffer a significant performance loss. Besides, for RPS and Without-RIS schemes, under different setups, it is unclear which one is better, but a narrow gap is provided. In general, when the number of D2D links is 6, the performance gap between the proposed algorithm and MPT, Without-RIS, RPS is $2.9\%$, $14.8\%$ and $18.7\%$, respectively.

\begin{figure}[t]
\begin{center}
\includegraphics*[width=0.8\columnwidth,height=2.2in]{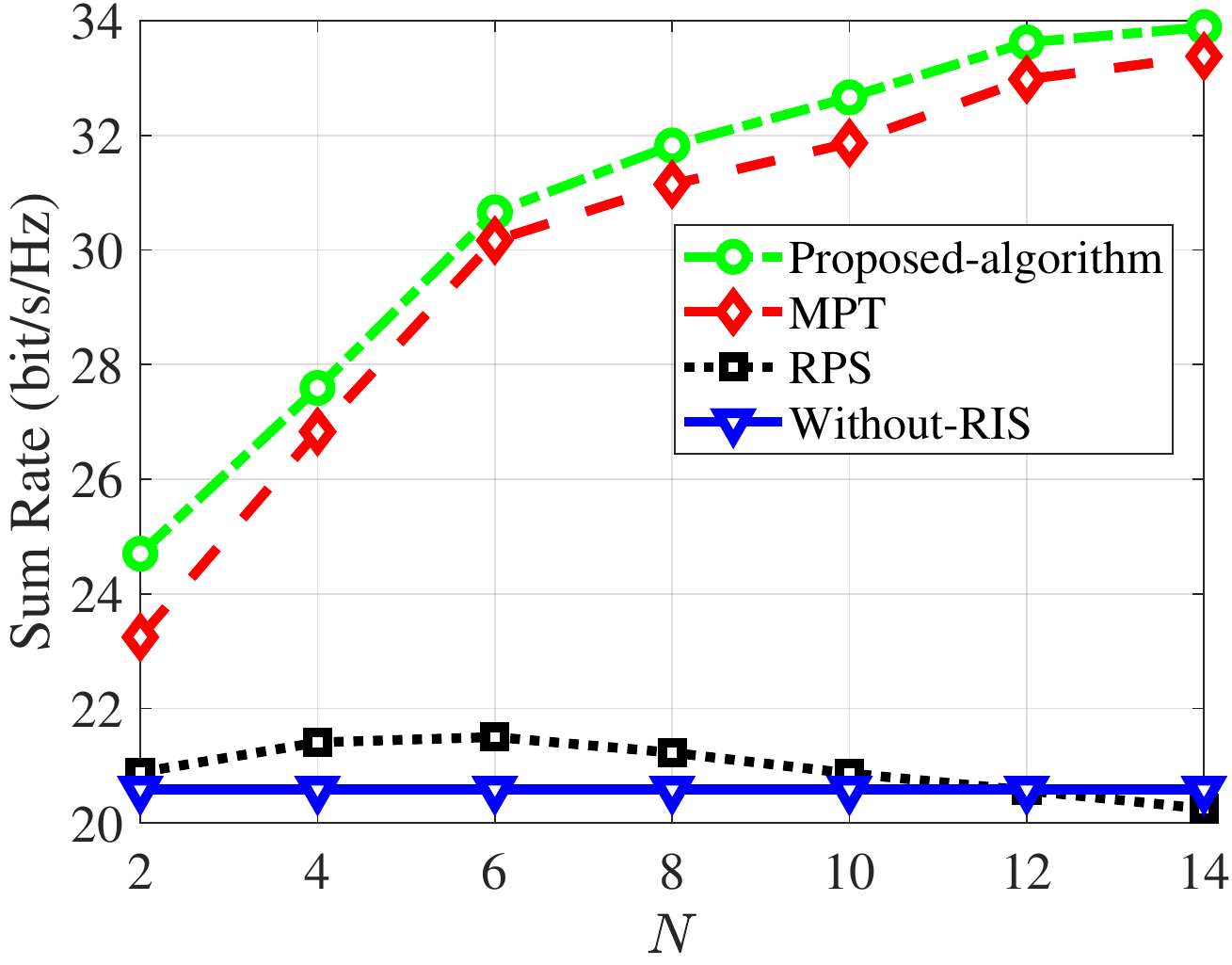}
\end{center}
\caption{Sum rate versus $N$.} \label{fig4}
\end{figure}

In Fig. \ref{fig4}, both the number of D2D links and $e$ are set to be 3. We plot the sum rates of four schemes versus $N$. As seen from the given results, the Proposed-algorithm outperforms others with the number of RIS elements varied from $2^{2}$ to $14^{2}$. Then, the proposed one and the MPT show an increasing trend as the number of RIS elements increases, which proves that by applying passive beamforming in the RIS-assisted D2D network, undesired channel interference can be more effectively suppressed. It worths pointing out that when the number of elements is large, the growth of these two schemes becomes slightly slower. This is because the RIS contributes an increase in the number of interference signal paths. Nevertheless, it is undeniable that continuing to increase the number of RIS elements and adaptively adjusting the phase shifts of elements is still considerable for these two schemes. However, for the RPS, without the constructive alignment of reflecting beams, the increase in the number of elements has caused the interference to increase significantly, which leads to a slower growth from 2 to 6, and results in a decreasing curve when $N$ exceeds 6. Finally, comparing the proposed algorithm with the Without-RIS scheme, a $64.6\%$ performance gain is obtained for $N=14$.

\begin{figure}[t]
\begin{center}
\includegraphics*[width=0.8\columnwidth,height=2.2in]{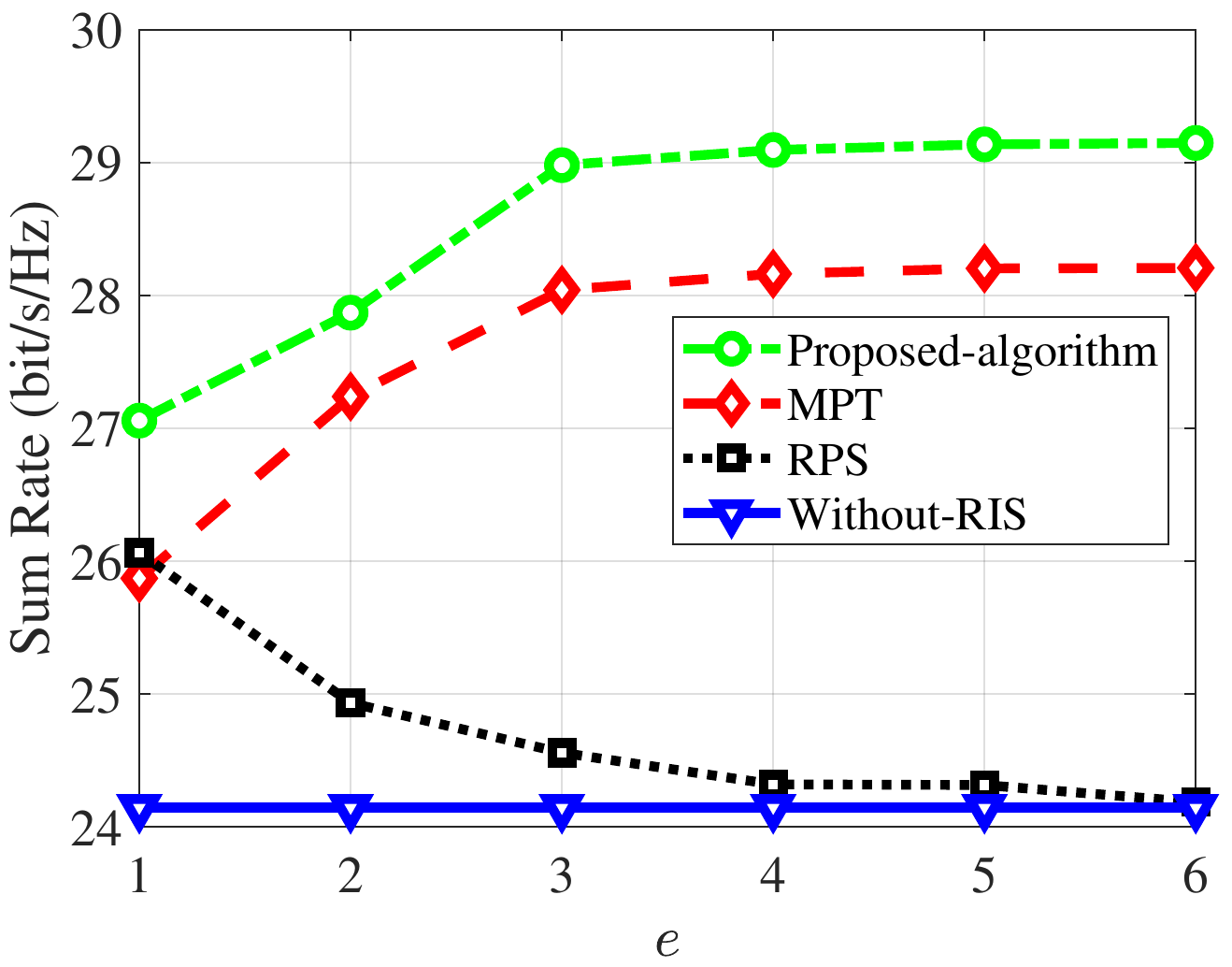}
\end{center}
\caption{Sum rate versus $e$.} \label{fig5}
\end{figure}
\begin{figure}[t]
\begin{center}
\includegraphics*[width=0.8\columnwidth,height=2.2in]{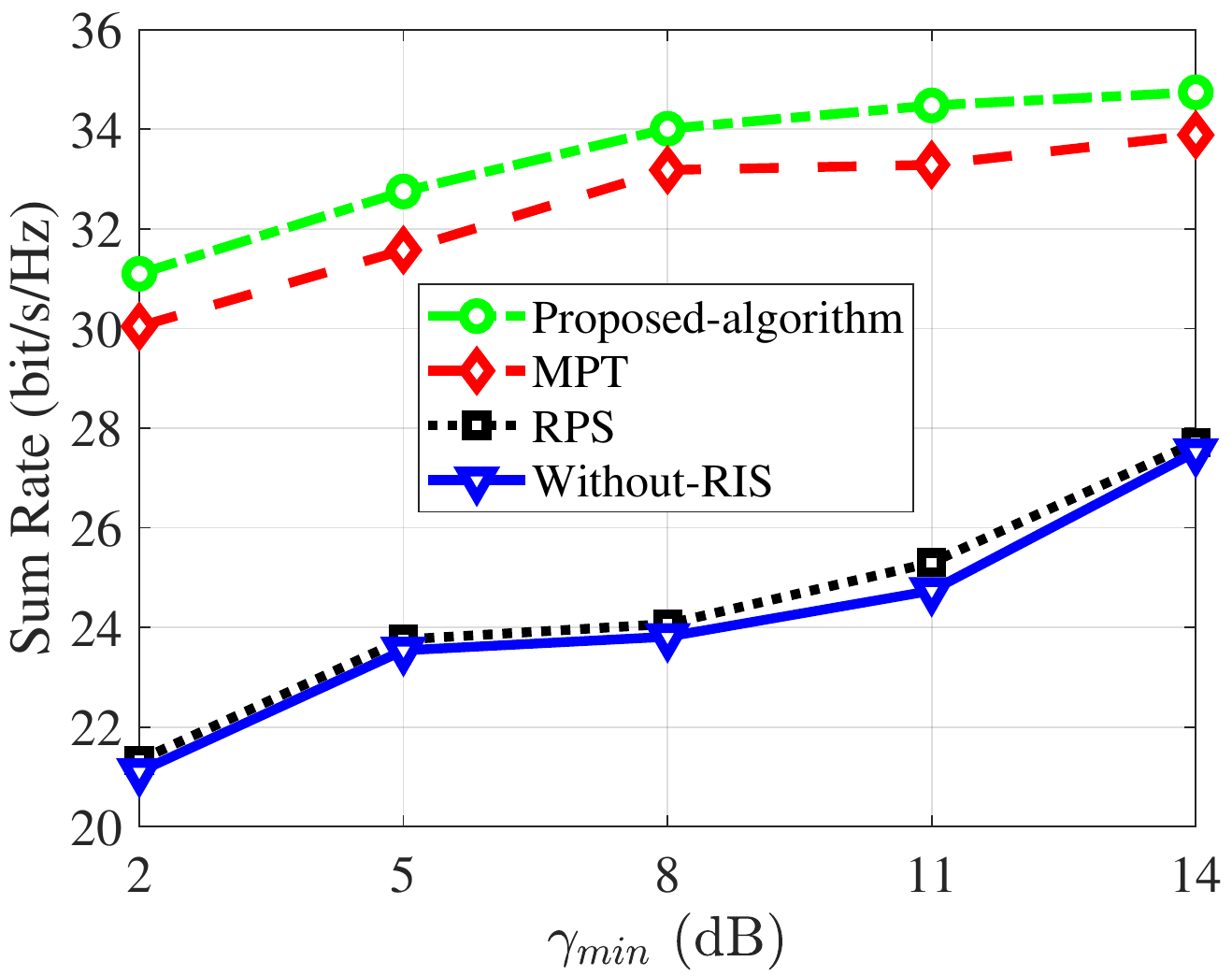}
\end{center}
\caption{Sum rate versus minimum SINR $\gamma_{min}$.} \label{fig6}
\end{figure}

In Fig. \ref{fig5}, $D=3$ and $N=3$. We plot the achievable sum rate performance while increasing the bit-quantization number from 1 to 6. There are several important observations. First, the advantage of the proposed algorithm is pronounced over the other three benchmark schemes. Second, the proposed and the MPT schemes gradually increase with $e$ grows from 1 to 4, and then basically remain unchanged from 5 to 6. This reveals that the system performance tends to be saturated when the number of quantization bits exceeds 4. In addition, the RPS exhibits a slight sum rate degradation. The reason behind this phenomenon is that the distance of the reflection link that generates the useful signal and the distance of the reflection link that generates the interference signal are both random, so the useful reflection signal is not necessarily stronger than the interference reflection signal. Under the parameter settings, the effect of changing $e$ on the superposition of multiple reflected interference links is more prominent.

\begin{figure}[t]
\begin{center}
\includegraphics*[width=0.8\columnwidth,height=2.2in]{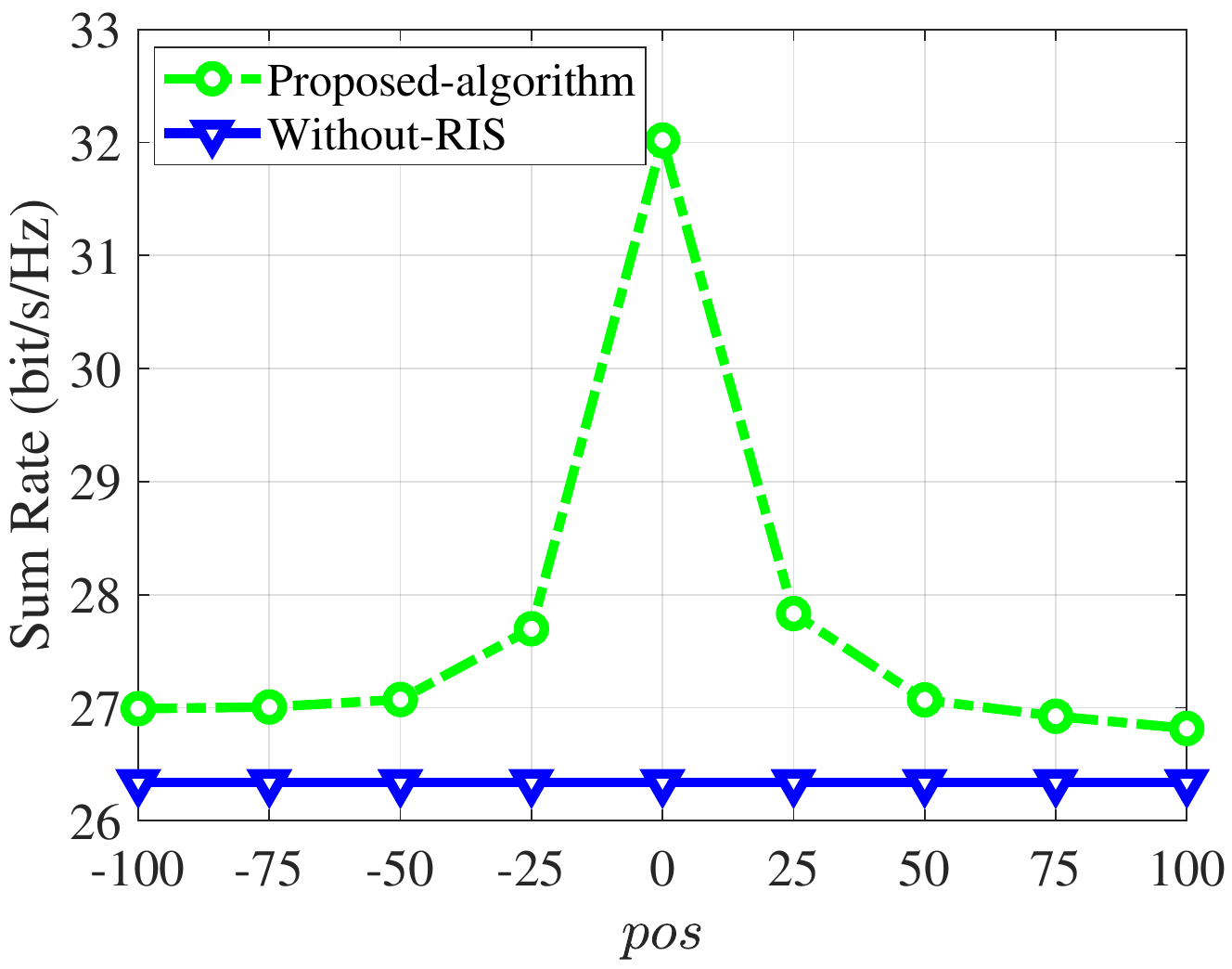}
\end{center}
\caption{Sum rate versus RIS position $pos$.} \label{fig7}
\end{figure}
\begin{figure}[t]
\begin{center}
\includegraphics*[width=0.8\columnwidth,height=2.2in]{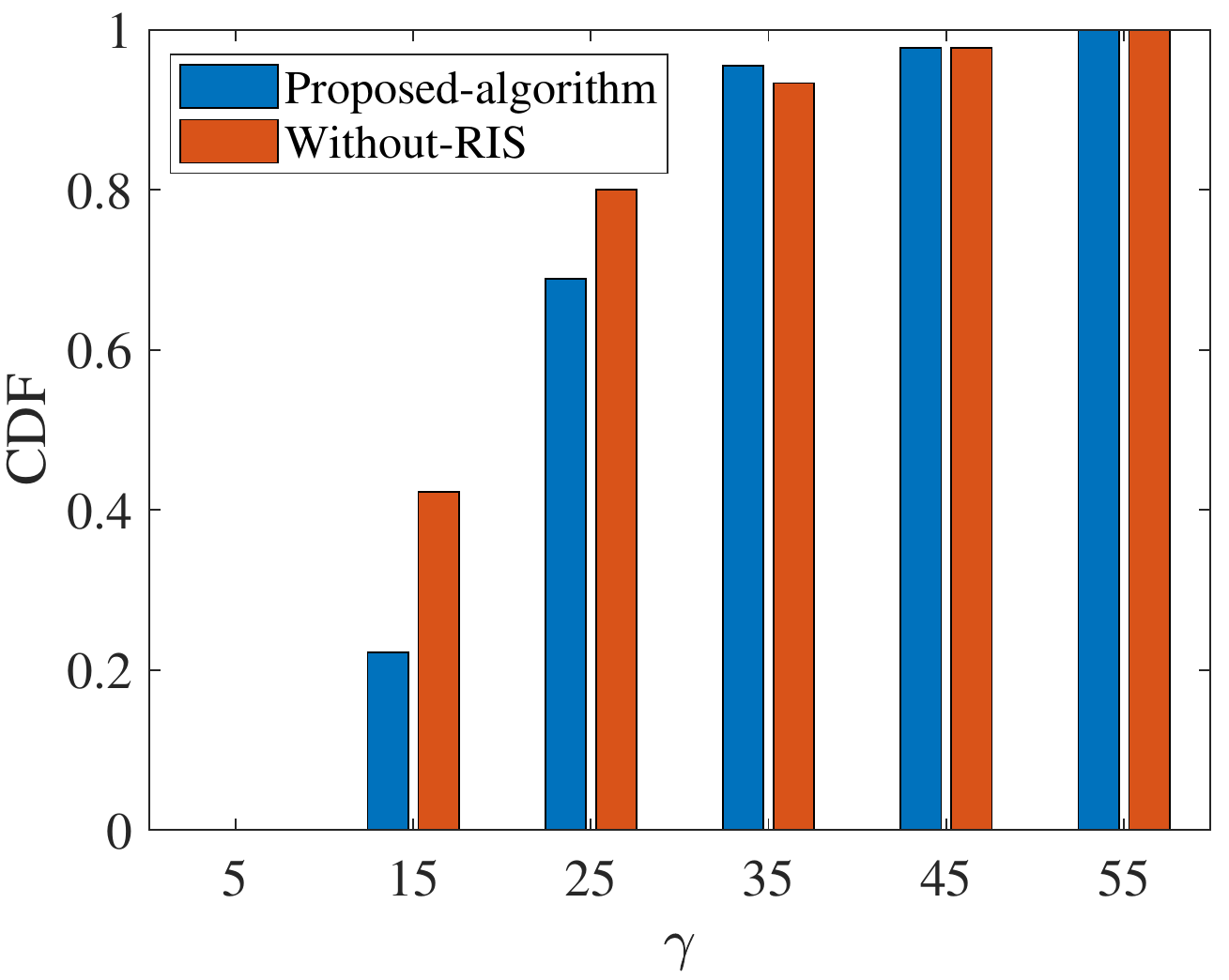}
\end{center}
\caption{CDF histograms for SINR $\gamma$.} \label{fig8}
\end{figure}

In Fig. \ref{fig6}, $D=3$, $N=4$ and $e=3$. We vary the $\gamma_{min}$ from 2 dB to 14 dB and depict the comparison of four schemes in terms of the sum rate. One can observe that, the curves of all schemes show an upward trend with the increase of the minimum value of individual SINR, and the proposed algorithm has a higher system sum rate than other schemes. There is no doubt that by improving the lower bound of the network QoS constraint, the performance of the system is also getting better. Furthermore, the schemes of RPS and Without-RIS have almost the same performance, and they are the worst two of the four schemes.

In Fig. \ref{fig7}, $D=3$, $N=3$ and $e=3$. We study the impact of the deployment location of RIS on the performance of the proposed algorithm. First, we choose nine positions as the vertices of the top left corner of RIS, which are $(0,-100,0)$, $(0,-75,0)$, $(0,-50,0)$, $(0,-25,0)$, $(0,0,0)$, $(0,25,0)$, $(0,50,0)$, $(0,75,0)$ and $(0,100,0)$, respectively. In other words, RIS is always in the $Y$-$Z$ plane and moves along the $Y$ axis. For convenience, we denote the position of RIS as $pos$, and different positions are distinguished by the distance between the top left corner vertex and the reference point $(0,0,0)$. Based on the above details, we plot the performance curves with the $pos$ from -100 to 100 for the proposed algorithm and the Without-RIS scheme. Interestingly, when the RIS position $pos$ is equal to -100 or 100, i.e., at the edge of the investigated area, the system performance is the worst. While $pos$ is equal to 0, which is the center of the long side of the rectangular area, the system performance is the best. This is owing to the fact that when the RIS moves closer to the center of the area, the distance of the reflection link will change accordingly, which results in an enhanced reflected signal and that the benefits of RIS are fully utilized. In general, the system performance is sensitive to the deployment location of RIS.

\begin{figure}[t]
\begin{center}
\includegraphics*[width=0.8\columnwidth,height=2.2in]{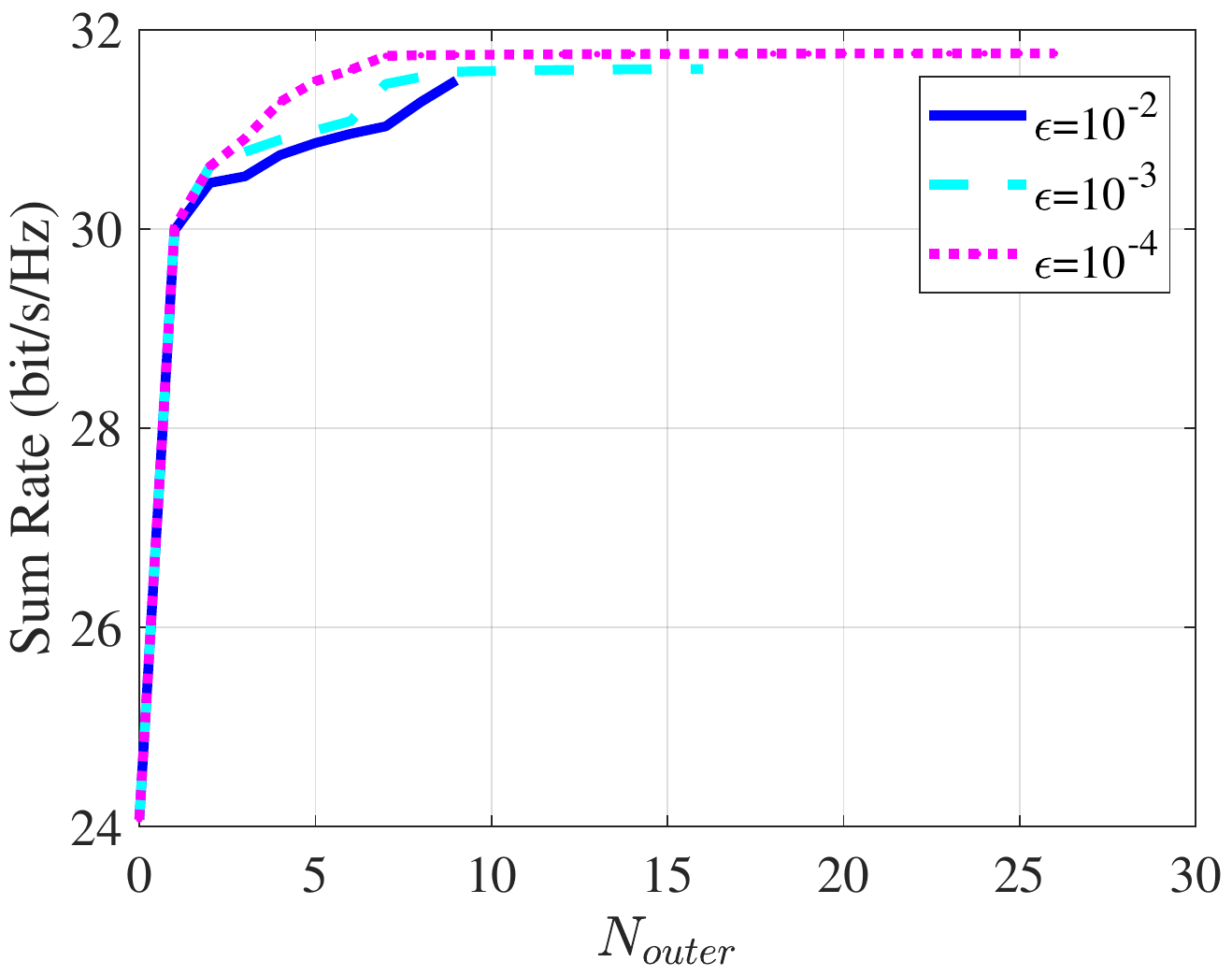}
\end{center}
\caption{Convergence behaviour of the proposed algorithm.} \label{fig9}
\end{figure}

In Fig. \ref{fig8}, $D=4$, $N=4$ and $e=3$. We plot the cumulative distribution function (CDF) of the SINR received by all links, which is denoted by $\gamma$, varying from 5 dB to 55 dB. Without loss of generality, all receivers are assumed to have the same SINR target, 5 dB. Comparing the proposed algorithm and the Without-RIS scheme, we can see that the performance gain brought by the deployment of RIS is completely non-negligible, and the effect of RIS on reducing interference of D2D networks is very considerable. This is because that, by adding more controllable reflection paths, RIS can boost the signal quality at the intended receiver.

In Fig. \ref{fig9}, $D=4$, $N=4$ and $e=3$. We simulate the complexity of the proposed algorithm under different stopping threshold settings, i.e., the number of iterations of the sum rate maximization algorithm. When the values of $\epsilon$ are $10^{-2}$, $10^{-3}$, and $10^{-4}$, respectively, the corresponding values of $N_{outer}$ required for convergence are 9, 16, and 26. First, we can see that, no matter the pre-defined threshold is large or small, the overall complexity of the sum rate maximization algorithm is still acceptable. In addition, the smaller the threshold, the slower the proposed algorithm reaches the stationary solution, but the obtained system performance is better.

\section{Conclusion}\label{S7}
In this paper, we have focused on uplink D2D-enabled cellular networks assisted by RISs. A joint power allocation design and RIS phase shift optimization problem have been studied to maximize the system sum rate under the individual QoS, power and practical discrete phase shift constraints. Due to the nature of mixed integer non-convex non-linear of the joint design problem, we have proposed an alternative optimization algorithm to obtain a sub-optimal solution. Finally, numerical results have validated the superiority of the proposed algorithm and we have observed that, by providing multiple controllable reflected signals based on the number of RIS elements, RIS can convert a channel in poor condition into a well-conditioned channel.

\bibliographystyle{IEEEtran}

\end{document}